\documentclass[preprint,showpacs,preprintnumbers,amsmath,amssymb]{revtex4}

\usepackage{graphicx}
\usepackage{amsmath}
\usepackage{subfigure}

\newcommand {\syninput} {\mathrm {input}}
\newcommand {\selfex }{\mathrm {selfex}}
\newcommand {\inh }   {\mathrm {inh}}
\newcommand {\ext }   {\mathrm {ext}}
\newcommand {\spike } {\mathrm {spike}}
\newcommand {\synasyn}{\mathrm {syn-asyn}}
\newcommand {\m }{\mathrm {m}}
\newcommand {\threshold}{\mathrm {th}}
\newcommand {\rest }{\mathrm {rest}}
\newcommand {\connect }{\mathrm {asm}}

\newcommand {\SYNCASYNC}{Syn-Asyn }

\newcommand {\patternfirst}{'$\times$' }
\newcommand {\patternsecond}{'$\star$' }

\begin{document}

\preprint{}
\title{
Effect of Synchronous Incoming Spikes on Activity Pattern in A Network of Spiking Neurons
}

\author{Takaaki Aoki}
 \affiliation{Department of Physics, Graduate School of Science, Kyoto University, Kyoto 606-8502, Japan }
\author{Toshio Aoyagi}%
 \affiliation{Department of Applied Analysis and Complex Dynamical Systems, Graduate School of Infomatics, Kyoto University, Kyoto 606-8501, Japan}

\date{\today}


\begin{abstract}
  Although recent neurophysiological experiments suggest that synchronous neural activity  is involved in  some perceptual and cognitive processes, the functional role of such coherent neuronal behavior is not well understood. As a first step in clarifying this role, we investigate how the temporal coherence of certain neuronal activity affects the activity pattern in a neural network. Using a simple network of leaky integrate-and-fire neurons, we study the effects of synchronized incoming spikes on the functioning  of two mechanisms typically used in model neural systems, winner-take-all competition and associative memory. We demonstrate that a pair of switches undergone by the incoming spikes, from asynchronous to synchronous and then back to asynchronous, triggers a transition of the network from  one state to another  state.  In the case of associative memory, for example, this switching controls the timing of the next recalling, whereas the firing rate pattern in the asynchronous state prepares the network for the next retrieval pattern.
\end{abstract}


\maketitle

\section{Introduction}

Synchronous neural activity, that is, coherent spiking behavior,  is  widely observed in various cortical areas,  and  is often stimulus-specific and task-specific \cite{Riehle1997,Gray89}. These observations  suggest that synchronous neural activity  may play a significant functional role in cognitive and perceptual processes \cite{Francisco2001,Engel01}. One proposed possibility is that such a coherent spiking is the signal for the binding of different attributes of an object into a single percept.
 To this hypothesis, we  have to first clarify the neural mechanism underlying this coherent activity. Over the last decade,  a considerable number of studies have been focused on determining  the dynamical mechanisms underlying synchronous firing in neural systems\cite{hansel95,Ermentrout96,vreeswijk94,Wang96,kopell00,Izhikevich00,Aoyagi02} , for example, elucidating the neural mechanisms that enable cortical neurons to synchronize and  how the neural network controls the degree of  synchrony in a stimulus-dependent manner.\cite{takekawa2003,Tiesinga2004} 
 The next important issue is to clarify how such  coherent activity contributes to information processing. However, there have been relatively few studies addressing this issue.\cite{Hopfield2003, Scarpetta2002} Considering the above situation, in this paper, we  explore the possible functional role of synchronized neural activity. 

Many electrophysiological experiments have shown that cortical neurons do not always synchronize and that the degree of synchrony is dynamically modulated by a number of factors, such as external stimuli and the internal state.  In fact, recent experiments show that the degree of coherence among spikes is modulated by attentional shifts.\cite{Riehle1997,fries2001}  Typically, it is observed that such coherent activity transiently lasts for a few hundred msec.  These facts indicate that the switching between asynchrony and synchrony provides a crucial signal for information processes in the brain. In developing a theory of information processing in the brain, it is necessary to understand how modulation of the level of synchrony influences the function of the neural network. In particular, we study the influence of synchrony-asynchrony switching in incoming spikes on the activity pattern on a neural network, resulting from two mechanisms typically employed in model  neural networks, winner-take-all competition and associative memory. 

\section{Winner-take-all Competition Model}
Let us begin our analysis by considering winner-take-all(WTA) competition, which is commonly used as a computational model to determine the best choice among different possible outputs of the network. Is is believed that this competition mechanism provides a fundamental mechanism used in the brain. For example, various models using this mechanism have successfully reproduced columnar structures found in the visual cortex \cite{amari1980,kohonen,Malsburg1973,winner-share-all}. Through WTA competition, the only neuron receiving the largest input in a network becomes active as the `winner'.  Recent simulations  have revealed that such  competition is extremely sensitive to the coherence among neuronal action potentials \cite{lumer}. We study this phenomenon theoretically and explore the possible role of synchronization in the functioning of neural networks. 

\begin{figure}
\begin{center}
  \includegraphics[width=7.5cm]{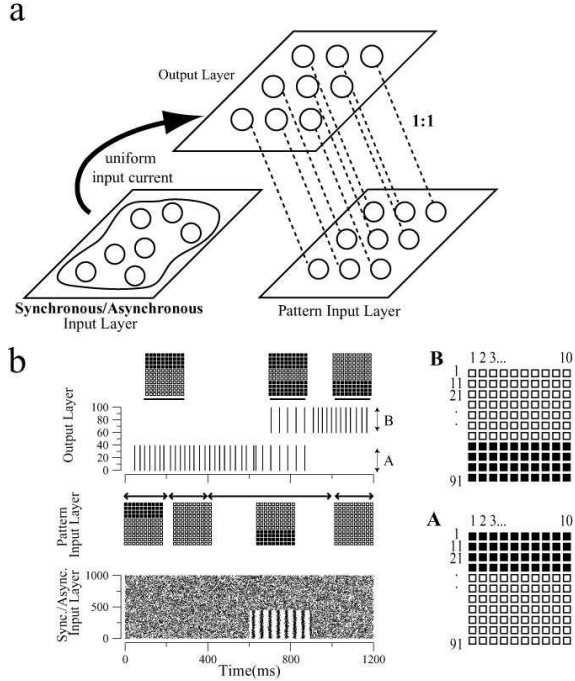}
\end{center}
\caption{(a)
  Schematic illustration of the winner-take-all(WTA) competition model. Each neuron   receives two external inputs, one from the asynchronous/synchronous input layer always uniformly inputs  and one from the pattern input layer neuron-specific inputs. (b) WTA competition suppressed by synchronized incoming spikes.  The upper two graphs display a firing pattern  and a rastergram of the output layer. The firing pattern of the output layer is shown as a 10x10-dot image, in which the gray level of a dot indicates the averaged firing rate over 200 ms (black under-bar), normalized by dividing by the maximum firing rate.  In the lower two graphs, the presentation of the pattern input layer and rastergram of the synchronous/asynchronous input layer are shown. Pattern input is presented as a 10x10-dot image in which a black dot indicates active state and a white dot indicates silent state, see detail right images. For an asynchronous input, the network exhibits typical behavior in the case of WTA competition. The winner pattern A is maintained even if a new pattern B is presented in the pattern input layer. However,  a brief synchronous input causes the WTA competition to reset, leading to a transition to a new winner pattern B.
}
\label{fig:wta_sim}
\end{figure}

A schematic of our WTA model is displayed in Figure \ref{fig:wta_sim}(a). As the single neuron model, we adopt the  leaky integrate-and-fire neuron model, in which the basic equation determining the membrane potential $V_i$ of the $i$-th neuron is given by
\begin{align}
  C_{\m} \frac{dV_i}{dt} &= - \frac{(V_i -V_{\rest})}{R_{\m}}+ I_{i}(t)  \\
 \tau_{\m} \frac{dV_i}{dt} &= - (V_i -V_{\rest})+ R_{\m}I_{i}(t)   \label{eq:ifmodel},
\end{align}
where  $R_{\m} = 100M\Omega$ and $\tau_{\m}= R_{\m}C_{\m} = 20ms$ are the membrane input resistance and  the membrane time constant, respectively.
Whenever $V_{i}(t) $ reaches the threshold voltage $V_{\threshold} = -50mV$,  a spike is generated, and $V_{i}(t) $ is instantaneously reset to the resting potential, $V_{\rest} = -65mV$. The total current $ I_{i}(t)$ consists of  the sum of all inputs to the neuron, given by
\begin{align}
 I_{i}(t) &= I_{i}^{\inh}(t)  + I_{i}^{\selfex}(t)  + I^{\synasyn}(t) + I_{i}^{\ext}(t),    \\
 \text{with} \quad     I_{i}^{\inh}(t)&= - \frac{g_{\inh}}{N} \sum_{j \not = i } \sum_{\spike}J( t - t_j -\Delta), \notag \\
    I_{i}^{\selfex}&= g_{\selfex} \sum_{\spike}J( t - t_i -\Delta), \notag
\end{align}
where $N$ is the total number of neurons, $t_i$ is the time at which the $i$-th neuron generates a spike, and $\Delta$ is the synaptic transmission delay. The postsynaptic potential is expressed simply  as  the alpha function $J(t)$, which is given by 
\begin{align}
 J(t) & =  \frac{t}{\tau^2} \exp(\frac{-t}{\tau}) \Theta(t), \notag
\end{align}
where $\tau$ is the time constant of a synapse, and $\Theta(t)$ is the step function, given by
\begin{align*}
 \Theta(t) & =  \begin{cases}
		 1 \qquad t \geq 0 \\
		 0 \qquad  t < 0 .
		\end{cases}
\end{align*}
The current $I_{i}^{\inh}(t)$ represents the effect of all-to-all uniform inhibitory couplings, which is  responsible for the WTA behavior of the network. We also assume that each neuron is self-connected by an excitatory synapse, $I_{i}^{\selfex}(t)$.  This single neuron in our model can be regarded as representing a population of the real neurons comprising a single columnar structure. Therefore, it is quite natural that such excitatory self-connections are introduced, corresponding to the mutual excitatory synaptic connections existing within an actual column of this kind. 

We assume that each neuron in the network receives input from two types of external sources. One is the synchronous/asynchronous (Syn-Asyn) input layer, in which the level of coherence among the input spikes can be controlled to examine the effect of the synchrony-asynchrony switching of incoming spikes on the activity pattern in a neural network. This uniformly applied current is generated according to 
\begin{align*}
 I^{\synasyn}(t) = \frac{g_{\syninput}}{N_{\syninput}}\sum_{\spike}J( t - t^{\synasyn} ),
\end{align*}
where $t^{\synasyn}$ is the time at which neurons receive a spike from a neuron in the \SYNCASYNC input layer and $N_{\syninput}$ is  the total number of neurons in this input layer.
 We consider the situation in which  this input layer possesses two modes, an asynchronous mode and a synchronous mode. In the asynchronous mode, each spike is  generated by a Poisson process. In the synchronous mode, 45 percent of the neurons in this layer fire periodically and synchronously  with spike timings distributed in a Gaussian distribution  of variance $\sigma = 4ms$. In both modes, we assume that the average firing rate is  25Hz.

 The other external source is the pattern input layer, which is  the neuron-specific current, given by
\begin{align*}
 I_{i}^{\ext}(t) &= I_i = \begin{cases}
			  \Delta I & \text{active state },\\
			   0        &\text{silent state}.
			  \end{cases}
\end{align*} 
Each neuron in this layer projects to one  neuron in the output layer in a one-to-one correspondence.
By controlling the activity of neurons in this layer, the biased current for each neuron in the output layer is determined, which affects the firing pattern selected through the WTA competition.

For these synaptic connections and external inputs, we use the parameter values $ \Delta = 1ms$, $\tau=4ms$, $g_{\inh} = 4.8 nS,g_{\selfex} = 3.0 nS$, $ g_{\syninput} = 7.14 nS$, $N_{\syninput} = 1000$ and $\Delta I = 7.5 pA$.

Figure \ref{fig:wta_sim}(b) depicts a typical effect of synchronized incoming spikes on the properties of the WTA competition. Here the simulation was performed with a network of 100 neurons. The system was started with the condition that the  \SYNCASYNC input layer is set in asynchronous mode, and in the pattern input layer, the currents $I_{i}^{\ext}(t)$ in the 1st through 40th neurons are active ( Pattern `A').
 Owing to the nature of the WTA competition,  only those neurons that receive a larger input are active (as seen during the first 200 ms in Fig.\ref{fig:wta_sim}(b)).
After these currents in the pattern input layer turn off, the firing of the winners(Pattern `A') is maintained.  Next, even though  another current pattern `B' applying to  some loser neurons(60-100th) is presented in the input layer, the firing pattern of winners remains stable (time steps 400ms - 600ms).  This result is typical behavior of WTA competition models. 


With the situation as described above, let us consider a transient synchronized spike inputs with temporal switching modes in the \SYNCASYNC input layer.  In this case, as shown in Figure \ref{fig:wta_sim}(b), the synchronized incoming spikes cause the old firing pattern to become unstable. Although the old firing pattern `A' is maintained  at first, the network eventually exhibits a  collective synchronized oscillatory state. When the asynchronous mode in the \SYNCASYNC input layer is recovered, a second firing pattern `B' is realized through WTA competition, as determined by the activity of the pattern input layer.
The other neurons(40-60th) that  receive no pattern input do not fire even in a synchronous input period.  It depends on the parameters whether or not  these neurons fire without pattern input during synchronous input period. However, even if these neuron fire, the next pattern `B' always is  realized in the following  asynchronous input period.
Therefore, it should be noted that a transition of  the firing pattern in the output layer is induced by synchrony-asynchrony switching in incoming spikes. 
This result means that the synchronized incoming spikes nullify the effect of the function of WTA competition, erasing the current pattern realized in the WTA competition, and thereby allowing it to make a transition to a new pattern prepared by the pattern input layer.

\section{analysis}
In order to understand  the result of the  numerical simulations described above, we now analyze the stability of the state selected through the WTA competition. %

For the convenience of analysis, first we rescale equation (\ref{eq:ifmodel}). Using a transformation given by
\begin{align*}
 s &= t / \tau_{\m} \\ 
 v_i &= \frac{V_i - V_{\rest}}{V_{\threshold} - V_{\rest}} ,
\end{align*}
equation \ref{eq:ifmodel} is rewritten as
\begin{align}
 \frac{dv_i}{ds} = -v_i +I_i (s),   \label{eq:ifanalysis}
\end{align}
where parameters are rescaled as
\begin{align*}
 \Delta &= \Delta / \tau_{\m}, \quad \tau = \tau /\tau_{\m} , \quad \Delta I =  \frac{\Delta I R_{\m}}{V_{\threshold} - V_{\rest}}, 
\quad  \text{and }   g_{x} = \frac{g_{x} R_{\m}}{ (V_{\threshold} - V_{\rest}) \tau_{\m}}. 
\end{align*}
%
%
Next, assuming that the total number of neurons in these layers is sufficiently large and the variance of the synchronous spike timing $\sigma = 0$, the form of the input currents can be simplified as
\begin{align*}
\qquad I_{i}^{\synasyn}(s) &= \begin{cases}
			g_{\syninput} /T_{0} \quad &\text{asynchronous}, \\
			g_{\syninput} ( \rho \sum_{n}J(s - nT_{0} )  + (1-\rho) /T_{0}) \quad &\text{synchronous,}
		       \end{cases} 
\end{align*}
where   $T_{0}$ is  the period of the input spikes from  \SYNCASYNC input layer. $\rho$ is the ratio of synchronized neurons. 

From the results of simulations,  we know that typically, winner neurons fire periodically and synchronously at the times $s_i = nT (n=1,2,...)$ for the asynchronous mode in the \SYNCASYNC input layer and $s_i = n(T_{0} - \theta)$ for the  synchronous mode.
Consequently, we consider only these solutions in this analysis and do not deal with other more complicated solutions. To analyze these solutions, the unknown parameters  $T$ and  $\theta$ should be determined in a self-consistent way.
In the asynchronous mode, integrating equation \ref{eq:ifanalysis} between $nT$ and $(n+1)T$, we get a self-consistent equation for the period $T$,
\begin{align}
 1 & =  (I_i + \frac{ g_{\syninput}}{T_{0}}) (1 - e^{-T})+ (g_{\selfex} - g'_{\inh}  ) K_{T}(-\Delta) \\
 &K_T(s) = e^{-T} \int_{0}^{T} e^{u} \sum_{m \in Z} J(s + u + mT) du,\quad g'_{inh} = g_{\inh} \frac{(N_{p} -1)}{N}
\end{align}
where  $N_{p}$ is the number of the winner neurons.
Solving this equation, $T$ is determined and the spike trains of winner neurons are determined.

Similarly, in synchronous mode, we also derive a self-consistent equation for $\theta$,
\begin{align}
 1  =  (I_i +\frac{ (1-\rho)g_{\syninput}}{T_{0}} )(1 - e^{-T_{0}}) + (g_{\selfex} - g'_{\inh} )K_{ T_{0}}(- \Delta) +  \rho g_{\syninput} K_{T_{0}}( \theta T_{0}),
\end{align}
and determine the spike trains of  winner neurons.

Given the spike trains of the winner neurons, the time evolution of the loser neurons can be calculated. 
Let $v_{i}^{n}(\phi)$  denote  the membrane potential of $i$-th neuron at the time $s = T(n + \phi) $, in which $\phi$ goes from $0$ to $1$.  The return map from $v_{i}^{n}$ to $v_{i}^{n+1}$ is given by
\begin{multline}
 v_{i}^{n+1}=\begin{cases}
    e^{-T} v_{\phi}^n + (I_i + \frac{  g_{\syninput}}{T_{0}})( 1 - e^{-T}) - g'_{\inh} K_T( \phi T - \Delta)  & \text{asynchronous,}\\
    e^{-T_{0}} v_{\phi}^n +  (I_i + \frac{  (1-\rho)g_{\syninput}}{T_{0}})( 1 - e^{-T_{0}})
  \\ \qquad   - g'_{\inh} K_{T_{0}}( (\phi - \theta )T_{0} - \Delta  )
		 + \rho g_{\syninput} K_{T_{0}} ( \phi T_{0})  &\text{synchronous.} 
		\end{cases}
\end{multline}
 $v_{i}^{n}(\phi)$  converges to the limit $v_{i}^{\infty}(\phi)$,
\begin{align}
 v_{i}^{\infty}(\phi)& = \begin{cases}
      \frac{ (I_i + \frac{ g_{\syninput}}{T_{0}})( 1 - e^{-T}) - g'_{\inh} K_T( \phi T - \Delta) }{1 - e^{-T}} & \text{asynchronous}\\
     \frac{ (I_i + \frac{(1-\rho ) g_{\syninput}}{T_{0}}( 1 - e^{-T_{0}})  - g'_{\inh} K_{T_{0}}( (\phi - \theta )T_{0} - \Delta  ) + \rho g_{\syninput} K_{T_{0}} ( \phi T_{0}) }{ 1 - e^{-T_{0}}  }&\text{synchronous} 
\end{cases}
\end{align}
The stability of the WTA state is ensured by the condition that none of the loser neurons fire, therefore,
\begin{align}
 v_{i}^{\infty}(\phi)<  1 \quad \text{$\phi$ in $[0,1)$} \quad \text{for all loser neurons $i$ } \label{eq:wta}.
\end{align}

\begin{figure}

\begin{center}
 \subfigure{
  \includegraphics[width=4cm]{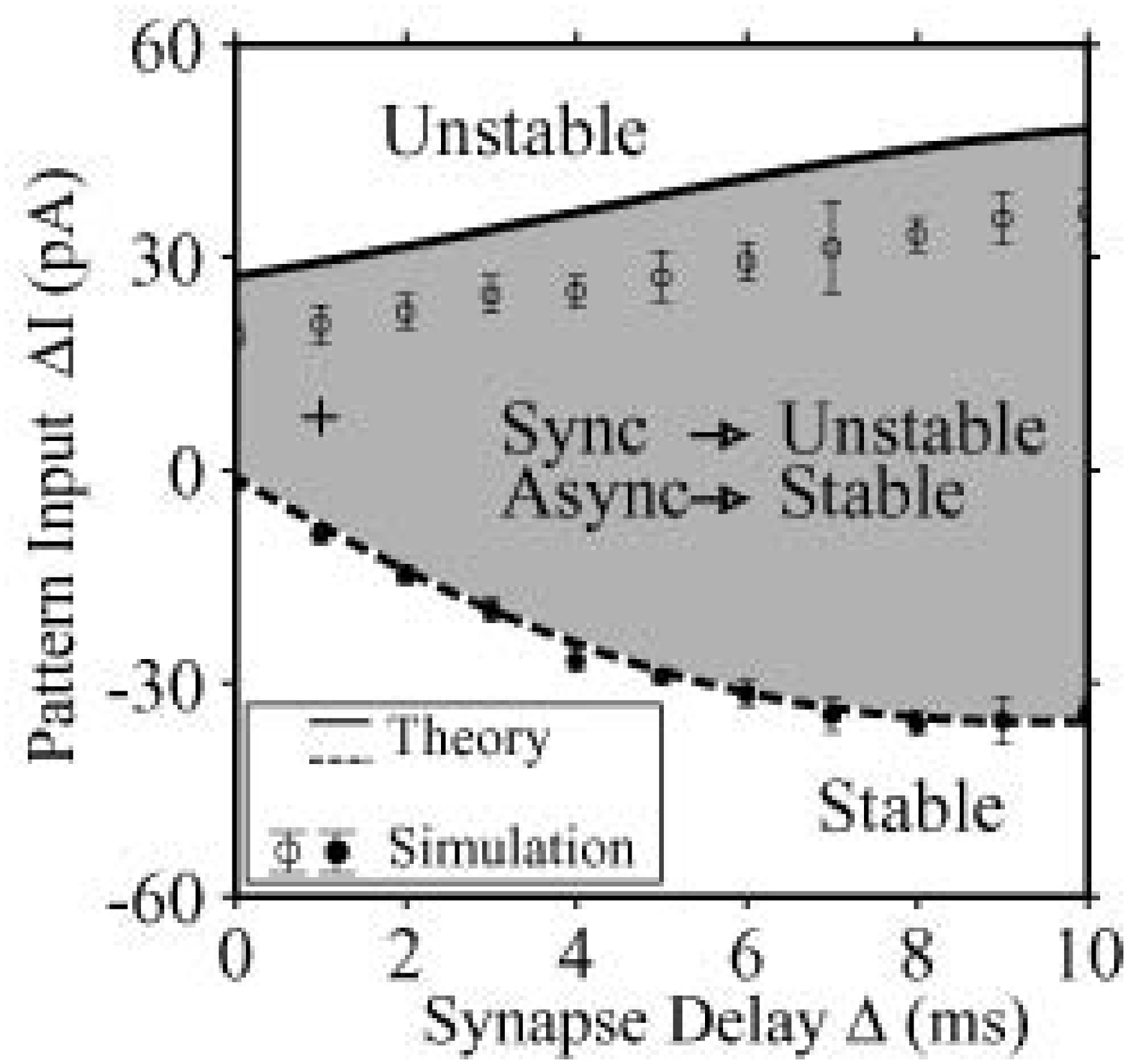}
  \label{fig:phase_diagram_inh}
 }
 \subfigure{
  \includegraphics[width=4cm]{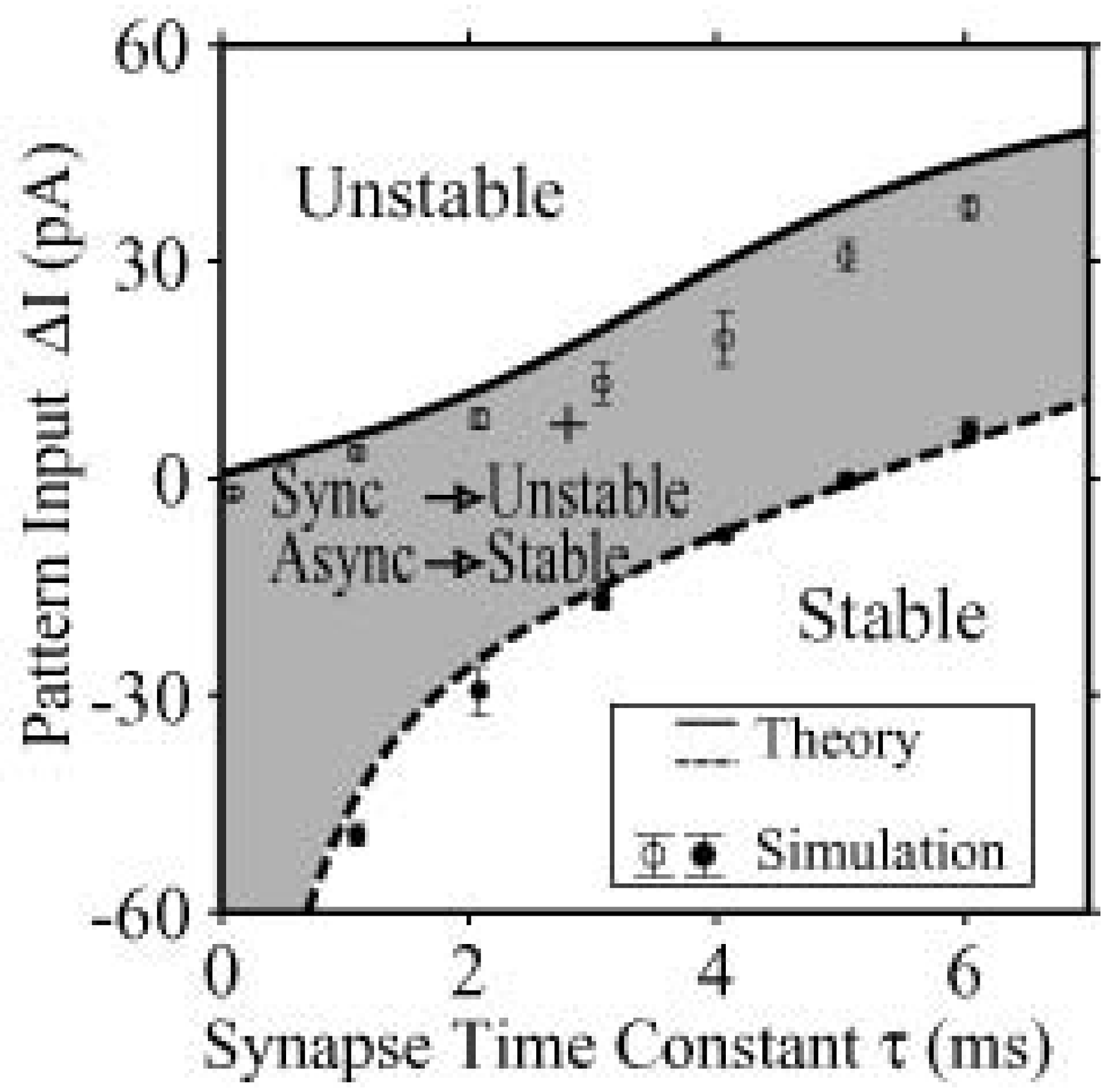}
  \label{fig:phase_diagram_selfex}
 }
\end{center}
 \caption{
   Two-parameter phase diagrams concerning the stability of the state selected through the winner-take-all competition.   This state is stable if the parameter value is located below the solid line (dashed line) for the case of  asynchronous (synchronous) incoming spikes. In the shaded regions, destabilization of the WTA state occurs in response to the switching from asynchronous to synchronous incoming spikes. The data points indicate simulation results. (a) Dependence of the stability on the synaptic transmission delay, $\Delta$, and the biased current from  pattern input layer, $\Delta I$. (b) Same as (a), except that the horizontal axis corresponds to the synaptic time constant, $\tau$. The parameter values used in Figure \ref{fig:wta_sim} is indicated by the $+$.
}
\label{fig:wta_phase_diagram}
\end{figure}
As depicted in Figure \ref{fig:wta_sim}(b) corresponding to $t = 400 - 600$ ms, we consider the situation in which the selected neurons remain active without a biased input and some loser neurons receive a biased current ($\Delta I$) from the pattern input layer. Using Eq.\ref{eq:wta}, in the situation described above, we determine  the condition under which the firing state of the selected neurons remains stable and the neurons receiving the biased input still cannot fire. The stable parameter regions obtained in this manner for the cases of the synchronous and asynchronous modes are displayed in Figure \ref{fig:wta_phase_diagram}.

In the shaded regions,  destabilization of the WTA state occurs in response to the switching of the mode of the \SYNCASYNC input layer from asynchronous to synchronous incoming spikes. This destabilization leads to a transition of the neuronal state into the next firing pattern, as determined by the pattern input layer.
Thus, if the parameter within the shaded region is used, the WTA network exhibits synchrony-induced switching behavior.
The validity of this analysis is shown by the numerical simulation data in Fig.\ref{fig:wta_phase_diagram}.
In addition, it also shows  that the result is not too sensitive to both the synaptic delay$\Delta$ and the synaptic time constant $\tau$.

\section{Associative Memory}
We next consider the case of the associative memory model \cite{hopfield1982,Gerstner1992}, in which information is stored in the synaptic weight matrix, $W_{ij}$. For the learning rule of the synaptic matrix, we adopt the Willshaw-type learning rule \cite{willshaw1969}, defined by 
\begin{align}
 W_{ij} = \Theta \left( \sum_{\mu = 1}^{P} \zeta_i^{(\mu)} \zeta_j^{(\mu)} \right), \label{eq:willshaw}
\end{align}
where  $P$ is the total number of the stored patterns, $\zeta_i^{(\mu)}$ represents the state in the $i$-th neuron in the $\mu$-th pattern and takes only two values 1 (firing) and 0 (quiescent).
Replacing the self-excitatory coupling in the WTA model with the synaptic matrix Eq.~\eqref{eq:willshaw}, the resultant dynamics for the $i$-th neuron are described by
\begin{align}
  \tau_{\m} \frac{dV_i}{dt} &= - (V_i -V_{\rest})+ I_{i}^{\inh}(t)  + I_{i}^{\connect}(t)  +I^{\synasyn} + I_{i}^{\ext}(t) , \label{eq:asm}\\
 I_{i}^{\connect} &= \frac{g_{\selfex}}{\bar{N}}\sum_{j \not = i} \sum_{\spike}W_{ij}J( t - t_{\spike}^j -\Delta) \notag
\end{align}
where the  factor $\bar{N} = \langle \sum_{j} W_{ij} \rangle_i$ is the average connections of synaptic matrix $W_{ij}$.

\begin{figure}
\begin{center}
  \includegraphics[width=8cm]{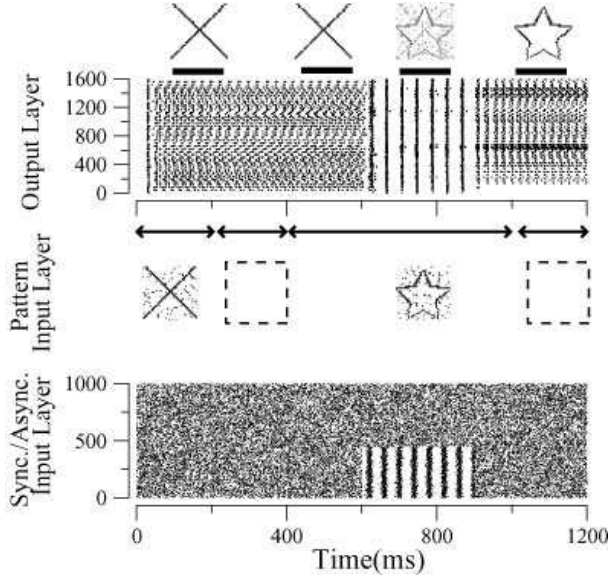}
\end{center}
 \caption{
  A possible role of synchronized incoming spikes in controlling the timing of the next retrieval in associative memory. The transition of the retrieval pattern is induced by the synchrony-asynchrony switching of the incoming spikes in the associative memory model. The upper two graphs display the pattern of the normalized firing rate presented as 40x40-dot images and a rastergram of the output layer.  The lower two graphs display the presentation of the pattern input layer (key pattern) and the rastergram of sync/async input layer.
}
\label{fig:assoiative_memory_sim}
\end{figure}

Figure \ref{fig:assoiative_memory_sim} depicts typical behavior of the associative memory model found in our numerical simulations, in which 10 patterns (\patternfirst, \patternsecond and 8 randomly generated patterns) are stored in the network of 1600 neurons. The `key' pattern is presented as the activity in the pattern input layer, displayed as a 40x40-dot image.  The `retrieval' pattern is displayed as a 40x40-dot image, in which  the  gray level of a dot indicates the averaged firing rate over 100 ms, normalized by dividing by the maximum firing rate. As shown in Figure \ref{fig:assoiative_memory_sim}, when a slightly distorted version of the pattern \patternfirst, consisting of \patternfirst  plus some noise, is presented as a key pattern, the network correctly realizes the pattern \patternfirst as the outcome of the associative memory functioning (0 - 200ms).
It is interesting that, if the spike trains in the \SYNCASYNC input layer synchronize, the existing retrieval pattern \patternfirst becomes unstable(600 - 900ms), leading eventually to the synchronous oscillatory state as in the case of the WTA model.  Then, when the input spike trains become uncorrelated again, the network comes to exhibit the new retrieval pattern \patternsecond in response to the key pattern presented by the pattern input layer(900 - 1200ms). Hence, the synchrony-asynchrony switching of the incoming spikes plays a crucial role in triggering a  change of the retrieval pattern, whereas the biased input from the pattern input layer prepares the network for the next retrieval pattern.

\section{discussions}
In the present paper, we have studied the manner in which the spike synchrony in the input layer can affect the activity pattern in a  neural network. We find that the  synchrony-asynchrony switching undergone by the incoming spikes plays a key role in triggering a transition from one state to another state, even if  the firing rate of incoming spikes is kept constant. For example, in the case of the associative memory model, the timing of the next retrieval can be controlled by such a  synchrony-asynchrony switching of the external uniform input spikes, whereas the averaged rate pattern in the pattern input layer  allows for the selection of the set of firing neurons for the next retrieval pattern. 

We will comment on a conventional way to reset activity patterns in WTA network. Because of the well-known multi-stability in a WTA network, the transition in the output layer can also be achieved  by changing the biased current from the pattern input layer. 
Therefore in Figure \ref{fig:wta_sim}(b), if the input current from the pattern input layer for pattern B is sufficiently large, a switching from A to B occurs.
This increase in the biased current corresponds to an increasing $\Delta I$ in Figure \ref{fig:wta_phase_diagram}, which makes the  WTA state  unstable.
By contrast with this conventional way, the important point of our model is that  the firing rate of incoming spikes does not change during a transition in the output layer.
The purpose of the present paper is to  propose an alternative mechanism to control the neuronal activity in a neural network by altering incoming spike synchrony only.

Some multi-recording experiments suggest that spike synchronization preferentially occurs in relation to a purely cognitive event, where firing rate modulations were absent \cite{Riehle1997,Lee2003}. 
For example, it is shown that before the onset of  voluntary movement, cortical neurons in the motor area transiently synchronize without changing the firing rate of each neuron.
In addition, many studies in the  recording of local field potential or electro-encephalographic activities also show that $\gamma$ frequency oscillation is observed in association with upcoming movements.
Therefore, some have  suggested that  these synchronous neural activities can be  regarded as the representation of an anticipation or a preparation of an upcoming movement.
In our understanding, such a synchronization in neural activity seems to play a role in triggering a transition to the next firing state for the upcoming movement.
Hence, we believe that some of the observed synchronous activities are essentially relevant to our result in the present paper.

\begin{figure}
\begin{center}
  \includegraphics[width=11cm]{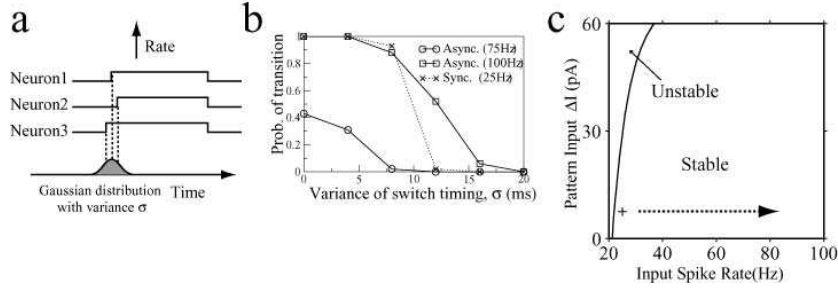}
\end{center}
 \caption{(a) Instead of synchronized spikes, a rate-modulated perturbation is applied to the neural network by changing the firing rate from 25Hz to 75 or 100Hz. The timing of the rate switching is distributed in a Gaussian distribution with variance $\sigma$. (b) The probability of a transition of the firing pattern as a function of  the fluctuation associated with the rate switching (solid curves). The dashed curves corresponds to the case of synchronized spikes. (c) Dependence of the stability of the state selected through the winner-take-all competition on the input spike rate in \SYNCASYNC input layer and the pattern input current $\Delta I$. Note that the constant rate for input is assumed.}
\label{fig:burst_input}
\end{figure}

Next, we will comment on the fact that the same transition can be achieved by uniform rate modulation in the \SYNCASYNC input layer. From a dynamical point of view, synchrony-asynchrony switching constitutes a kind of perturbation leading to a transition from one attractor to another. It can be imagined that the same effect could be achieved through some other kinds of perturbations that  do not involve synchrony among spikes, such as uniform modulation of the firing rate. To clarify this point, we investigated the effectiveness of such rate-modulated spikes numerically. In this case, instead of the synchronized incoming spikes, we used  an abrupt change in the firing rate from 25Hz to 75Hz or 100Hz, but the timing of the change in firing rate is distributed according to  a Gaussian distribution with variance $\sigma$(Figure \ref{fig:burst_input}a).
Figure \ref{fig:burst_input}b displays the probability of a successful transition induced by such a  modulation of the firing rate as a function of the width of the variance $\sigma$.
We see that a change to 75Hz is insufficient to achieve  a high probability of transition, even when the changes in firing rates are completely synchronous (i.e. $\sigma$=0). In the case that the firing rate is increased to 100Hz and $\sigma$ is sufficiently small, the probability for a transition is comparable to the case of synchronized spikes (see Sync.25Hz in Fig.\ref{fig:burst_input}b). Therefore, we conclude that a large and  simultaneous increase in firing rate is required to realize a high probability for a transition. This implies that this modulation can practically be regarded as  a kind of temporal coding.

Finally, we will point out that the synchrony-induced transition has different properties from the rate-modulation-induced one as mentioned above.
As shown in Fig.\ref{fig:burst_input}(c), the WTA state is  still stable even with a  high firing rate.
This analysis is based on the assumption that the incoming spike rate is constant.
Therefore, a sudden increase in the firing rate may cause a transition to another firing pattern.
In fact, the numerical simulations show that  a large and almost simultaneous increase in firing rate over all neurons can trigger a transition in the output layer.
%
On the other hand, by switching to a synchronous spike input, the WTA state becomes unstable. Therefore, the old firing pattern of the WTA state is eventually broken for the several synchronous spikes, whereas the old pattern is still maintained for  the first synchronous spike (Fig.\ref{fig:wta_sim}(b)).  This result indicate that  synchrony-induced transition mechanism  is different from  the rate-modulation-induced method.
We believe that instead of  rate-modulated type of perturbations, synchrony-asynchrony switching in incoming spikes  may provide another sophisticated way to cause a transition of the firing pattern.



%

\section*{Acknowledgments}
We thank T.Fukai for preliminary discussions and introducing us to this topic.
We are supported by the Japanese Grant-in-Aid for Science Research Fund from the Ministry of Education, Science and Culture.

\newpage 


\end{document}